\documentclass[reprint, amsmath,amssymb, aps]{revtex4-1}

\usepackage{graphicx}
\usepackage{epstopdf}
\usepackage{dcolumn}
\usepackage{bm}

\begin{document}

\preprint{APS/123-QED}

\title{Liquid state of hydrogen bond network in ice}

\author{M. I. Ryzhkin$^1$, A. V. Klyuev$^2$,  V. V. Sinitsyn$^1$ and I. A. Ryzhkin$^1$}

\email{ryzhkin@issp.ac.ru}

\affiliation{$^1$Institute of Solid State Physics RAS, Chernogolovka, Moscow District, Russia\\
$^2$National Research Lobachevsky State University of Nizhni Novgorod, Nizhny Novgorod, Russia}

\date{\today}

\begin{abstract}
Here we show that the Coulomb interaction between violations of the Bernal-Fowler rules leads to a temperature induced step-wise increase in their concentration by 6-7 orders of magnitude. This first-order phase transition is accompanied by commensurable decrease in the relaxation time and can be interpreted as melting of the hydrogen bond network. The new phase with the melted hydrogen lattice and survived oxygen one is unstable in the bulk of ice, and further drastic increase in the concentrations of oxygen interstitials and vacancies accomplishes the ice melting. The fraction of broken hydrogen bonds immediately after the melting is about 0.07 of their total number that implies an essential conservation of oxygen lattice in water. \\
\begin{description}
\item[PACS numbers]

64.70.Dv, 05.50.+q, 61.20.Lc
\end{description}
\end{abstract}

\maketitle

{\sl Introduction.} - For many reasons liquid water and its most widespread solid phase, called hexagonal ice, are among the most important and interesting substances. The phase transition between them is at the heart of many processes occurring under natural Earth conditions, and is of a great importance for all living beings and for numerous applications. The specific physical properties of water and ice are largely due to the crucial difference in their oxygen and proton lattices. 

In the hexagonal modification of ice (for brevity further we will use the term "ice" for it) oxygen ions form an ordered hexagonal lattice similar to the lattice of hexagonal diamond. Herewith each proton has two possible positions on each hydrogen bond shifted from its center, thus the number of possible positions is twice the number of protons. In the ground state of ice, the protons are distributed over these positions according to the ice or the Bernal-Fowler rules  \cite{Bernal1933}. The ice  rules state: (a) there are two and only two protons near each oxygen ion, and (b) there is one and only one proton on each hydrogen bond. Fig.1 shows a fragment of hexagonal ice lattice. 

Pauling showed that the number of proton configurations satisfying the ice rules is exponentially large and assumed that they all have the same energy, which implied the existence of a non-zero entropy per molecule at zero temperature \cite{Pauling1935}. This conclusion contradicts to the generally accepted statement about the zero entropy at zero temperature known as the third law of thermodynamics. Nevertheless, the  Pauling hypothesis is confirmed by numerous direct and indirect experiments comprehensively reviewed in \cite{Petrenko1999}. The theoretical justification of Pauling's hypothesis and explanation of the fact why the Coulomb interaction does not disrupt degeneracy of ground state configurations  were given much later \cite{Gingras2001, Isakov2005}.

\begin{figure}
\includegraphics[width=0.48\textwidth]{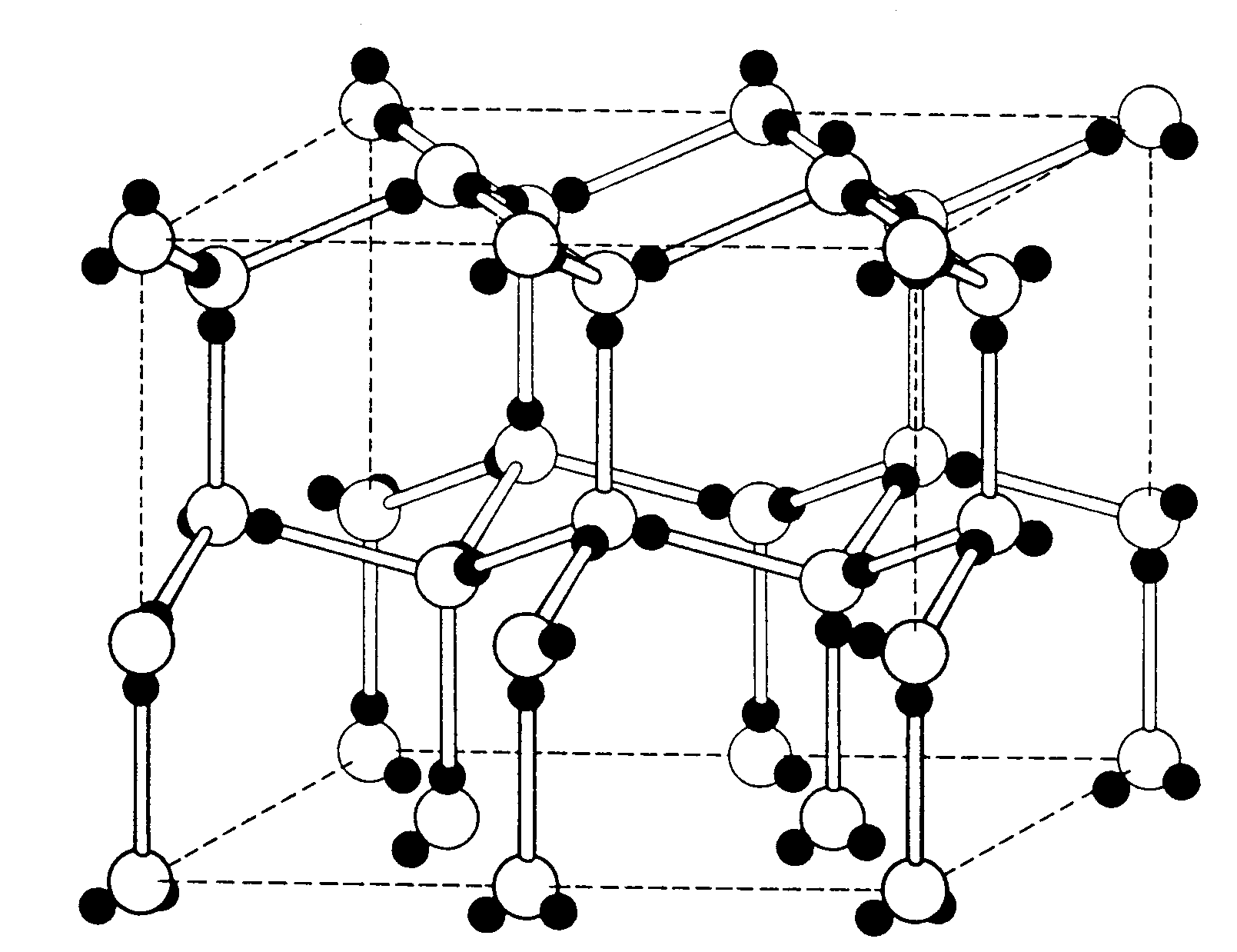}
\caption{Fragment of hexagonal ice lattice. Oxygen ions are shown by light circles, they form a hexagonal wurtzite lattice. Protons are shown by dark circles, their distribution over hydrogen bonds satisfies the ice rules.}
\label{fig.1}
\end{figure}

Any proton displacement in a ground state configuration leads to the violation of the ice rules and to the increase in energy (see Fig. 1). Hence, relaxation of the proton lattice is forbidden at zero temperature. However, it becomes feasible at finite temperatures due to violations of the ice rules called proton defects. The ionic defects  ($H_3O^+, OH^-$) and the bond defects ($D, L$) have the lowest activation energies equal to $1.40$eV and $0.68$eV per a pair of ionic and bond defects, respectively (see Fig. 2,3 on the next page). The proton  defects are also characterized by effective charges, mobilities and equilibrium concentrations. In fact, they are classical quasi-particles, which provide an economical description of the proton lattice response to various applied disturbances \cite{Bjerrum1951, Granicher1958, Jaccard1964}. Indeed, using them one has moved from considering a concentrated collection of protons to a more dilute array of itinerant and weakly interacting quasi-particles.

\begin{figure}
\includegraphics[width=0.5\textwidth]{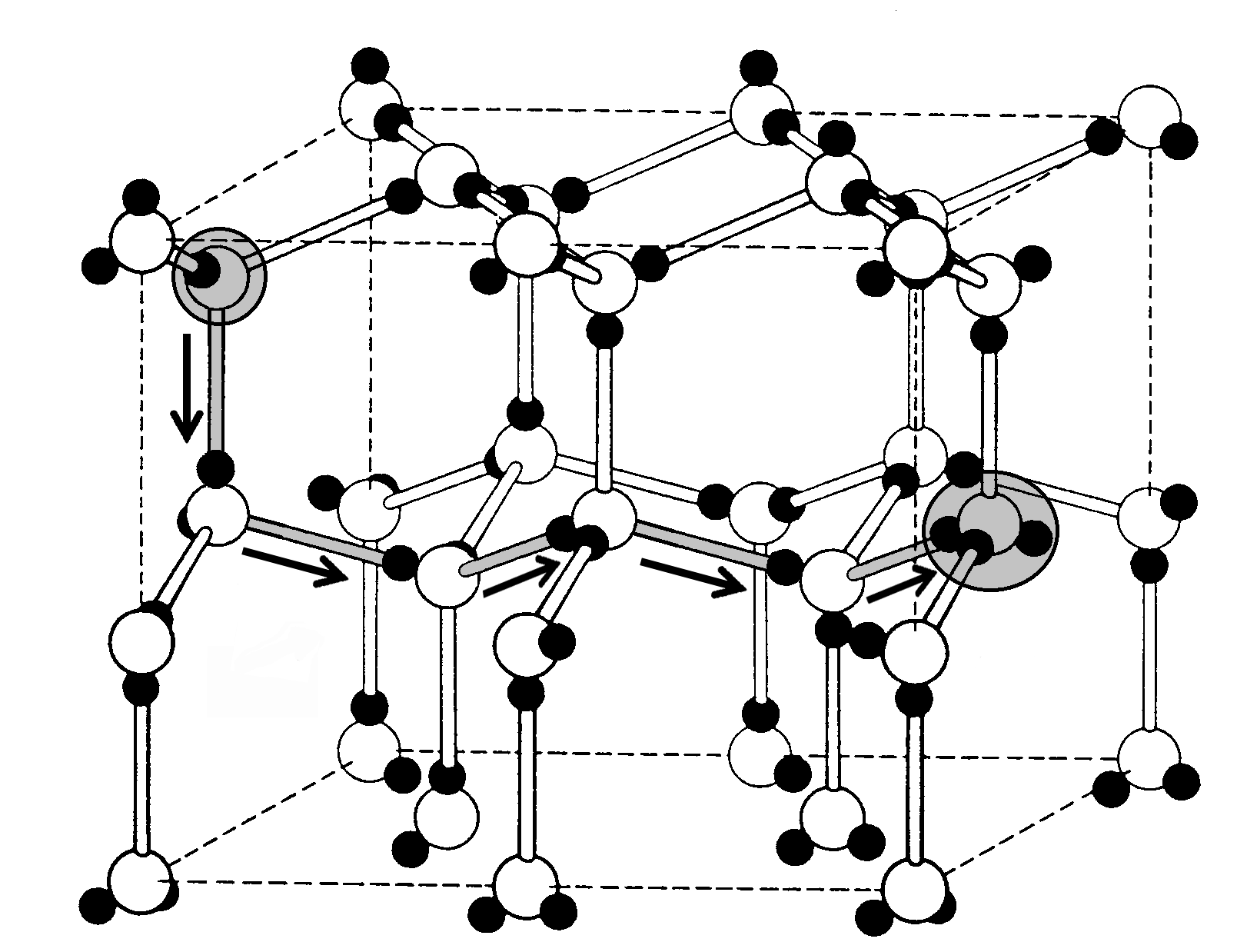}
\caption{Creation and removal of ionic defects: $H_3O^+, OH^-$ defects are shown by large grey circles (right and left circles, respectively).}
\label{fig.2}
\end{figure}

With regard to water, the most of researches examine the structural models and pair distribution function $g_{OO}(r)$, which can be obtained from experiments on neutron and x-ray scattering. All structural models of water can be divided into two large groups. The first group includes two-component models and originates from Roentgen's work \cite{Roentgen1892}, in which water was considered as a mixture of an ice-like component and a denser one similar to "normal liquid". In subsequent years, Roentgen's ideas were substantially developed in \cite{Nemethy1962, Hagler1972, Jhon1976}. Samoilov's interstitial model \cite{Samoilov1965} and Pauling's clathrate one \cite{Pauling1959} would also be included into this group.\\ 
\indent The second group includes homogeneous models and originates from the Bernal-Fowler work[ \cite{Bernal1933}, where they assumed that water has a crystalline oxygen lattice similar to that of ordinary ice. The only difference is that the oxygen lattice of water is a combination of two types of lattice: a less dense tridimite-like lattice and  a denser quartz-like one. Later, this model was improved in \cite{Pople1951, Bernal1964, Angell1971, Sceats1979}. The second group also include the percolation model of Stanley and Teixeira \cite{Stanley1980}. Note that all these models imply some (static or dynamic) preservation of the oxygen lattice of ice in water.\\ 
\indent The purpose of our paper is to study a new feasible state of ice with melted proton lattice, but with comletely preserved oxygen one. The relaxation times of proton system in this state are very short, and it can be called as a liquid state of hydrogen bond network.

\begin{figure}
\includegraphics[width=0.5\textwidth]{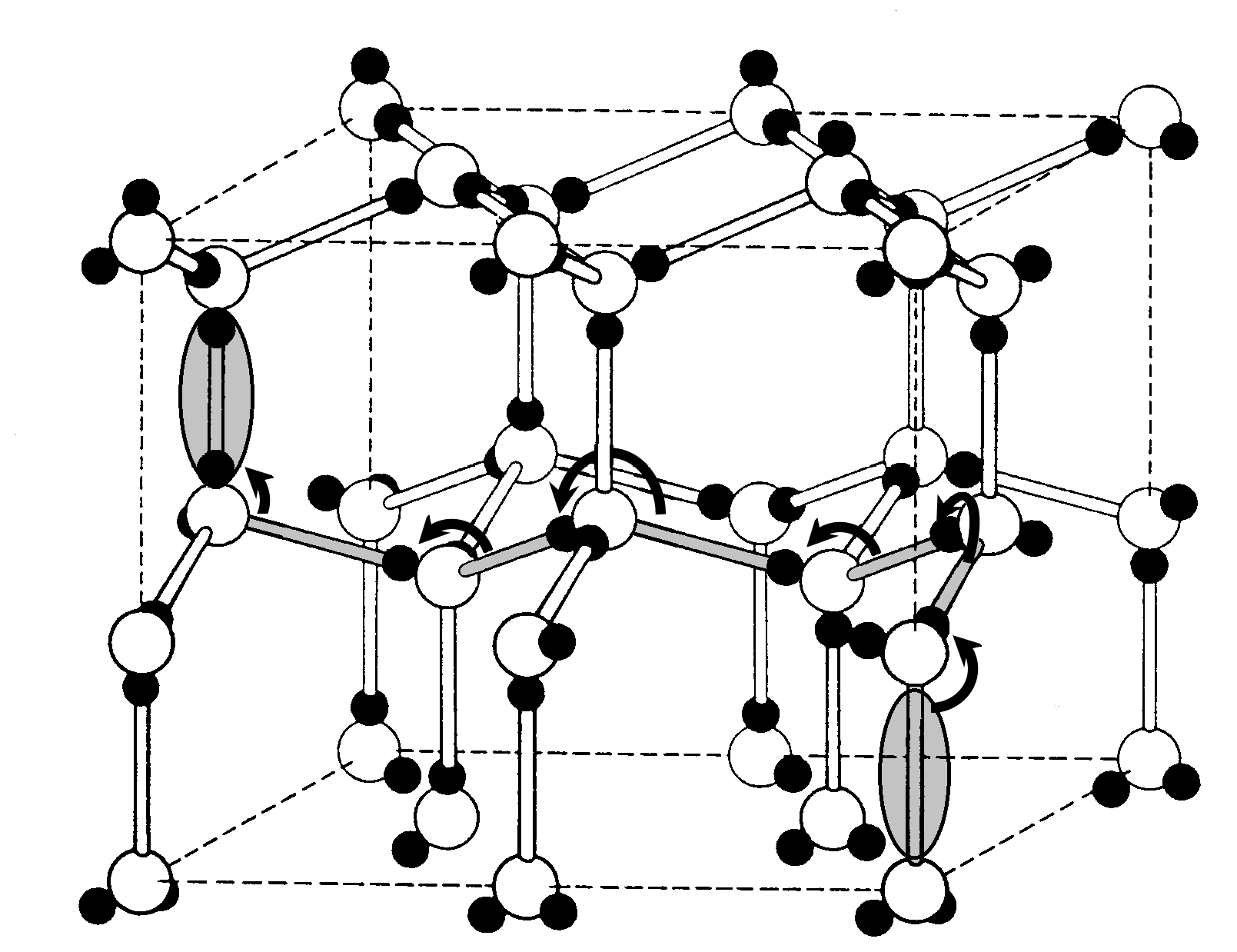}
\caption{Creation and removal of bond defects or Bjerrum defects:  $D, L$ defects are shown by large grey ovals (left and right ovals, respectively).}
\label{fig.3}
\end{figure}

{\sl Model and basic equation.} - Our model consists in the following. As mentioned above, the proton distribution in ice is determined by the ice rules, but at non-zero temperatures their violations or proton defects should arise. With a fixed activation energy of defects $E$  their concentration increases exponentially with growth of temperature as  $n\propto \exp(-E/2kT)$.  However, this description does not take into account the Coulomb interaction between charged defects. As we will show later, the Coulomb interaction enables a decrease in the activation energies of defects with the increase in their concentrations. The decrease in activation energies can  lead to a step-wise growth in defect concentrations at some critical temperature, which in turn makes the oxygen lattice instable and accomplishes the ice melting.

Let us proceed to the detailed description of our model and consider the formation process of a defect pair. For definiteness let it be a pair of bond defects. We divide the creation of free defects into two stages (see Fig. 3). At the first stage, a proton moves from hydrogen bond to the nearest one that produces a pair of charged bond defects at the minimum distance $b=\sqrt{2/3}r_{OO}$, where  $r_{OO}=2.76\cdot10^{-8}$cm is the distance between adjacent oxygen sites. This process requires the energy $E^1_{34}$,  where the low index  $34$ designates a pair of  $D,L$ defects. At the second stage, the charged defects should be removed from each other at a sufficiently long distance, which is about the mean inter-defect distance. This process requires the energy  $E^2_{34}$, which can be estimated as follows. Taking into account the Debye screening, we write the potential $\varphi_{34}(r)$ of a  $D$-defect with the size $b$ at the distance  $r>b$ in the following form \cite{Robinson1959}

\begin{equation}
\varphi_{34}(r)=\frac{q_{34}\exp(\kappa b)}{1+\kappa b}\cdot\frac{\exp(-\kappa r)}{r}
\label{eq:1}
\end{equation}

where the inverse screening length is determined by both ionic and bond defects

\begin{equation}
\kappa=\sqrt{\frac{8\pi}{kT}(q^2_{12}n_{12}+q^2_{34}n_{34})}
\label{eq:2}
\end{equation}

Here $n_{12},n_{34}$  are the volume concentrations of ionic and bond defect pairs,  $q_{12},-q_{12},q_{34},-q_{34}$ are the effective charges of  $H_3O^+,OH^-,D,L$ defects, respectively. Then, the second part of the activation energy per a bond defect pair is equal to

\begin{equation}
E^2_{34}=-\int\limits_{b}^{\infty}q_{34}\frac{d\varphi_{34}}{dr}dr=\frac{q^2_{34}}{b}\frac{1}{1+\kappa b}
\label{eq:3}
\end{equation}

The total activation energy of a bond defect pair equals $E_{34}=E^1_{34}+E^2_{34}$.  Similar reasoning leads to the same concentration dependence of the activation energy for ionic defects with replacement in minimum distance $b \to a=r_{OO}$.

It is essential that the second part of the energy $E^2_{34}$ is a function of the inverse screening length and it tends to zero in the limit  $\kappa \to \infty$. Hence, an increase in the concentration of defects enhances the screening and leads to the decrease in the energy  $E^2_{34}$ that increases the defect concentration. At a sufficiently high temperature such a "positive feedback" can generate a step-wise growth of the defect concentration.

To study this effect, one has to derive the free energy of proton system $F=E-TS$ as a function of defect concentrations, and then explore its minima. The entropy of ice proton system as a function of defect concentrations was calculated in \cite{Petrenko1997}. Using its results and aforecited ones for energy, we get the free energy of proton system per a water molecule in the form
\begin{eqnarray}
f=E^1_{12}x+2E^1_{34}y+\frac{q^2_{12}/a}{1+\kappa a}+\frac{2q^2_{34}/b}{1+\kappa b}\nonumber \\
 +kT\Big[2x\ln{x}+(1-2x)\ln{(1-2x)/3}\Big]\\
+2kT\Big[2y\ln{2y}+(1-2y)\ln{(1-2y)}\Big] \nonumber
\label{eq:4}
\end{eqnarray}
Here  $x=n_{12}/N, y=n_{34}/2N$ are relative concentrations of defect pairs,  $N=3\sqrt{3}/8r^3_{OO}$ is the number of water molecules per unit volume. The first two terms are due to the formation energies of defect pairs at the first stages, the third and the forth terms are the energies required for realizing the second stages of defect formation. The minimum distances between defects, i.e. the distances resulting from the first stages, equal $a$  and $b$ for ionic and bond defects, respectively. The fifth and sixth terms represent the contribution of entropy to the free energy. The equilibrium values of concentrations $x,y$  correspond to the absolute minimum of free energy.

But before study the free energy, one has to introduce the high frequency dielectric permittivity of ice into Eq. (4). For the distances exceeding by far large typical interatomic distances, this permittivity is $\epsilon_{\infty}(r\gg r_{OO})=3.2$. If the distances between charges are much shorter than the interatomic ones, then the high frequency permittivity should be $\epsilon_{\infty}(r\ll r_{OO})=1$. However, Eq. (4) involves the interaction between charges at intermediate distances $a,b\approx1$, where the permittivity is not well defined. For this reason, we consider the values $\epsilon_{\infty}(a), \epsilon_{\infty}(b)$  as adjustable parameters with values taken from the interval from $1$ to $3.2$. In addition, we take the experimental values for effective charges $q_{12}=0.62e, q_{34}=0.38e$, where $e$ being the proton charge, and the following values for the adjustable paramenters: $E^1_{12}=0.64$eV,  $E^1_{34}=0.05$eV, $\epsilon_{\infty}(a)\approx 2.6527$, and $\epsilon_{\infty}(b)\approx 1.4548$. Then, using the Eqs. (2,4), we come to the expression for the free energy as a function of $x,y$, and its numerical analisys shows the following results.

{\sl Results and discussion.} - First, we have found that $y\gg x$ at all temperatures, and this is a consequence of the difference in activation energies: $E^1_{12}> E^1_{34}$. At typical temperatures the concentration of ionic defects $x$ is about $10^{-6}$ of bond defect concentration. Hence, the most of results can be obtained assuming  $x \approx 0$.\\
\begin{figure}
\includegraphics[width=0.52\textwidth]{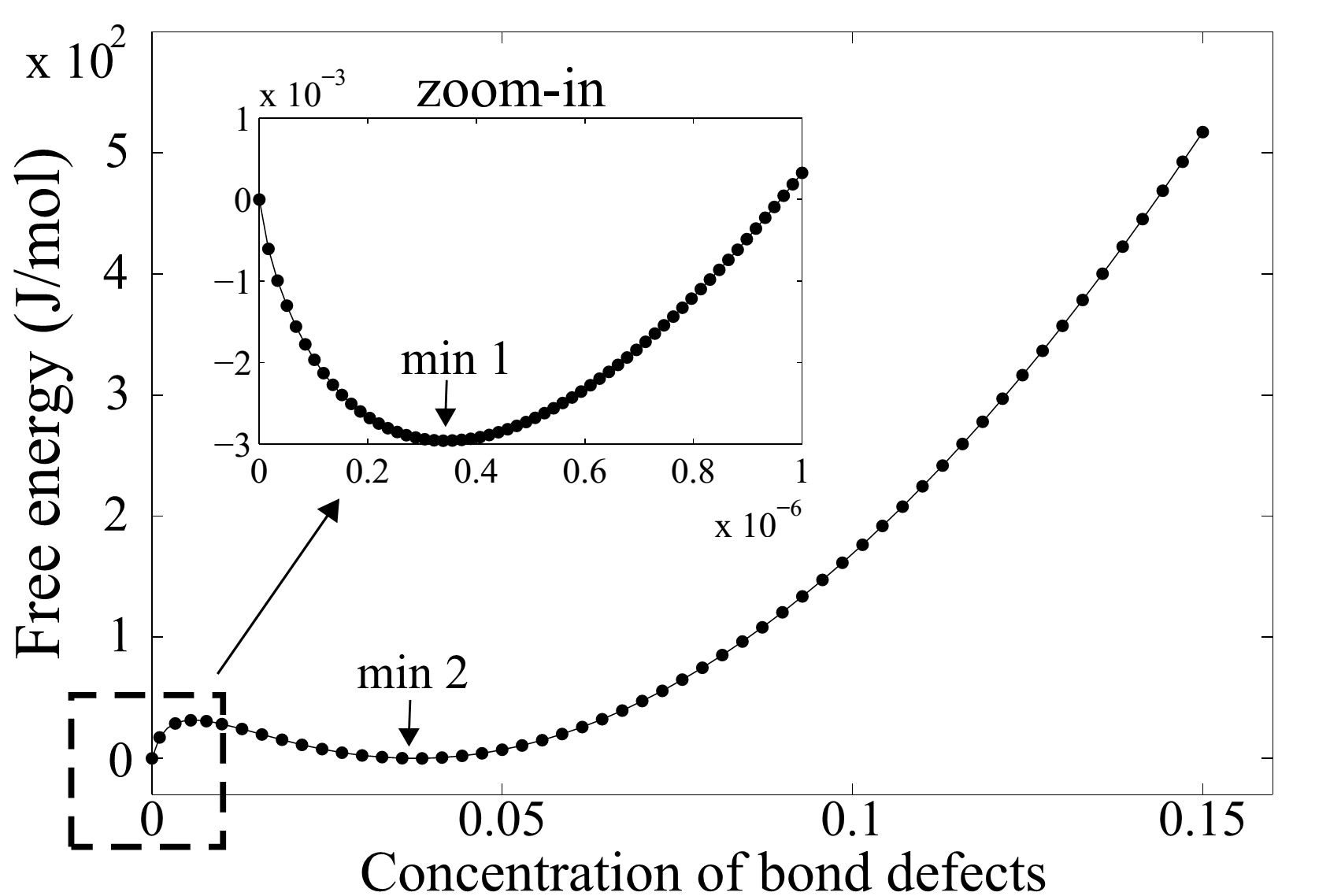}
\caption{The free energy of proton system of ice per a molecule as a function of the bond defect concentration at $T_c=273$K.}
\label{fig.4}
\end{figure}
\indent Second, we have revealed that with the chosen values of parameters the free energy has two local minima at  $y=y_1$ and $y=y_2$, where $y_1\ll y_2$.   At low temperatures, namely at $T<T_c$, the first minimum at $y=y_1$ provides a lower value of the free energy and therefore corresponds to the stable state, which is the common ice with low concentrations of bond defects. The second local minimum at $y=y_2$  corresponds to a metastable state in this case.  As the temperature increases, the first minimum rises, whereas the second one lowers, and at temperature $T=T_c$   they become equal. This corresponds to the first-order phase transition of ice into a new phase, which is the ice with a liquid state of hydrogen bond network. Fig. 4 presents the graph of free energy with two equal minima at $T=T_c$. Since the minima occur at very different values of $y$, it is impossible to display them on the same graph. That is why the left minimum is displayed in an enlarged scale on the inset. Practically exact coincidence of the critical temperature and the melting one is a consequence of the parameter choice.

\begin{figure}
\includegraphics[width=0.492\textwidth]{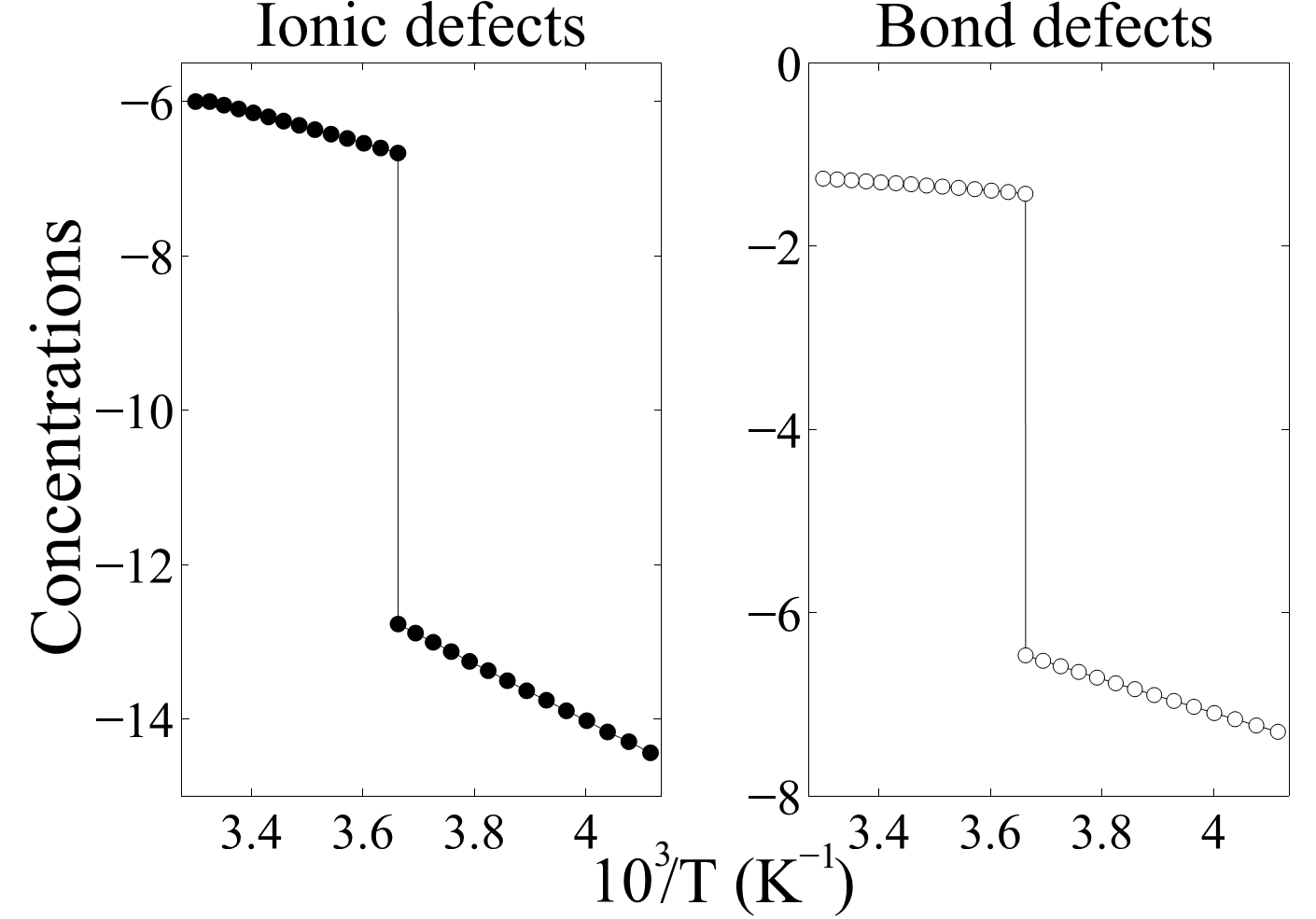}
\caption{Logarithms of relative concentrations of proton defects as a function of inverse temperature.}
\label{fig.5}
\end{figure}

We interpret the new phase with the drastically increased concentration of bond defects, but with still preserved oxygen one as an intermediate phase  between ice and water. Experiment shows that such a phase is never realized in the bulk samples of ice, since the increase in bond defect concentrations leads to a disorder in the oxygen lattice. However, it could exist near ice surface, in ice/water in nanotubes or at high pressure. The intermediate phase is a precursor of the real melting of ice, so it can be useful for the understanding of ice melting. Some properties of the intermediate phase can survive even after the destruction of oxygen lattice, we can observe them in properties of water at temperatures close to the melting point. For example, the higher density of water can be explained by the increased concentration of bond defects and relaxation of oxygen lattice near them, for example, by complexes of $D$-defect with oxygen interstitial as suggested in \cite{Haas1962}.\\
\indent Third, Fig.5 presents equilibrium defect concentrations as functions of inverse temperature. The concentrations is seen to exhibit a rather exact exponential dependence on inverse temperature like  $x,y \sim \exp(-E_{x,y}/2kT)$ both below and above the critical temperature. The activation energies for ionic defects are equal to  $1.40$eV and  $0.74$eV for ice and water, respectively, while for bond defects they are $0.68$eV and $0.14$eV. These values are very close to the most reliable experimental results \cite{Petrenko1999, Artemov2014, Klyuev2014, Artemov2015}. At the critical temperature, the defect concentrations undergo a step-wise increase by 6-7 orders of magnitude.  Just after the phase transition, the bond defect concentration is about 0.07.\\
\begin{figure}
\includegraphics[width=0.47\textwidth]{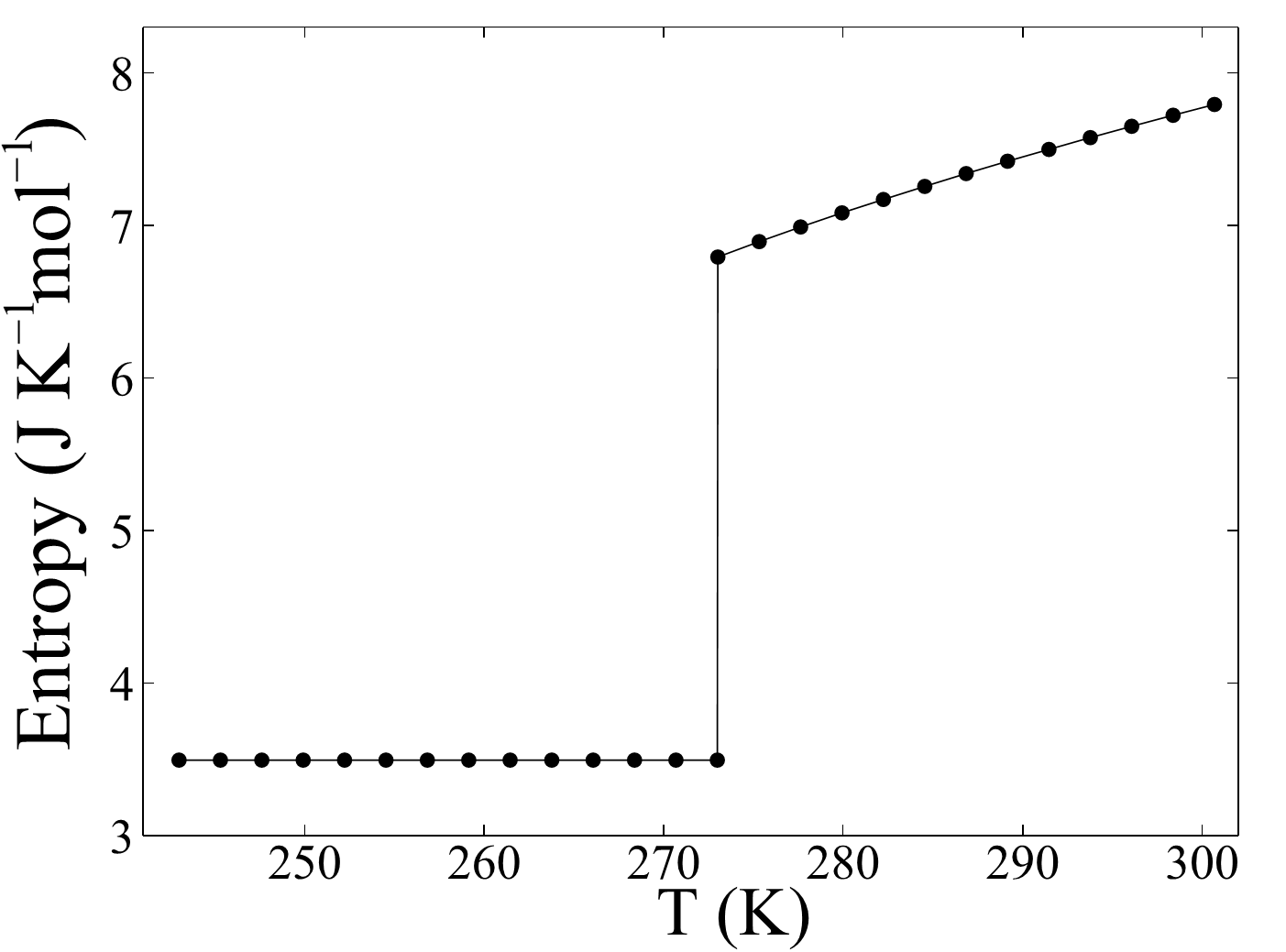}
\caption{Configuration entropy of proton system as a function of temperature.}
\label{fig.6}
\end{figure}
\indent Fourth, Fig.6 shows the temperature dependence of entropy that exhibits a jump at the critical temperature. The value of entropy jump provides the latent heat of transition which is equal  to approximately 0.17 of the experimental value for real ice.  However, one should add the contribution from the destruction of oxygen lattice. We can estimate this contribution using the two-component model of amorphous ice proposed in \cite{Ponyatovsky1994}. In accordance with \cite{Ponyatovsky1994}l, breaking the hydrogen bonds produces two components: low density clusters of the common hexagonal ice and high density clusters, i.e., the regions with a collapsed oxygen lattice. The change in entropy accompanying the transition contributes about 0.30 of the ice melting heat. Hence, the total contribution of the proton and oxygen lattices to the ice melting heat reaches approximately 0.47. The remaining difference, about 0.53 of ice melting heat, can be attributed to the vibrational degrees of freedom which were omitted in our approach.\\
\indent At last, we note that our model agrees with study of electrical properties ice and water near the melting point \cite{Artemov2014, Klyuev2014, Artemov2015}. Indeed, as it was shown above, the relative concentration of the bond defects does not exceed 0.07. Thus, only about 0.07 of oxygen lattice in the new phase could be destructed because of the oxygen lattice relaxation. Nevertheless, this concentration of oxygen defects can be sufficient for the oxygen lattice to acquire some properties of liquid \cite{Frenkel1946}. But the most of oxygen lattice can be described by the Jaccard theory \cite{Jaccard1964}. At the first glance, the high frequency (determined by bond defects) and low frequency (determined by ion defects) conductivities should increase in the same proportion. However, the experiment shows that the high frequency conductivity really increases by 6-7 orders of magnitude, whereas the low frequency conductivity increases only in 30-40. This can be explain by an essential decrease of mobilities of ionic defects. Indeed, the ionic defects move via jumps of proton along hydrogen bonds, and the bonds with $D$ and $L$ defects are dead-ended for their movement.\\

\begin{acknowledgments}
This work was supported by the Program of Presidium RAS P11 "The matter at high pressure", by Ministry of Education and Science of Russian Federation: the State Task "Carrying out of research works", project code 2183 and Agreement ¹ 02.Â.49.21.0003 with Lobachevsky State University of Nizhni Novgorod.
\end{acknowledgments}

\bibliography{liquid}

\end{document}